# Implied correlation from VaR[1]

# John Cotter[2] and François Longin[3]

[1] The first author acknowledges financial support from a Smurfit School of Business research grant and was developed whilst he was visiting ESSEC Graduate Business School.

[2] Director of Centre for Financial Markets, Department of Banking and Finance, Smurfit School of Business, University College Dublin, Blackrock, Co. Dublin, Ireland, Tel.: +353-1-7168900. E-mail: john.cotter@ucd.ie.

[3] Professor of Finance, Department of Finance - ESSEC Graduate Business School - Avenue Bernard Hirsch, B.P. 105 - 95021 Cergy-Pontoise Cedex - France. E-mail: flongin@essec.fr.

# Implied correlation from VaR

**Abstract**

Value at risk (VaR) is a risk measure that has been widely implemented by financial institutions. This paper measures the correlation among asset price changes implied from VaR calculation. Empirical results using US and UK equity indexes show that implied correlation is not constant but tends to be higher for events in the left tails (crashes) than in the right tails (booms).

JEL Classification: G12

Keywords: Implied Correlation, Value at Risk

# 1. Introduction

Value at risk measures the potential loss of a market position over a given time-period and for a given confidence level. For example, a 1-day 99% VaR of $ 1 000 000 means that over the next trading day, in one case over one hundred, the portfolio loss will be higher than $ 1 000 000. In order to compute the VaR of a portfolio, it is necessary to make assumptions on the distribution of asset price changes (see Jorion, 2000; and Dowd, 2002; for a presentation of VaR methods). The Gaussian distribution with a constant correlation matrix is often assumed in practice. This paper takes the reverse approach by inferring the correlation implied by VaR calculation.

# 2. VaR calculation

In this section we consider a portfolio composed of two assets, asset 1 in proportion $x_1$ and asset 2 in proportion $x_2$ ($x_1$+ $x_2$=100%). The frequency used to measure the asset and portfolio price changes, which also corresponds to the holding period used to compute the VaR, is denoted by $f$. The portfolio VaR is computed by two approaches: first, by considering the distribution of asset price changes of the whole portfolio and then compute the portfolio VaR as a quantile of this distribution; second, by considering the distribution of price changes of each asset, then compute the VaR for each asset and finally compute the portfolio VaR by using an aggregation formula.

**a) The portfolio approach**

In the portfolio approach, a time-series of the portfolio price changes for a given frequency $f$ is built from the time-series of each asset price: $\Delta P_{port} = x_1 \cdot \Delta P_1 + x_2 \cdot \Delta P_2$. The distribution of portfolio price changes is then built in order to compute the portfolio VaR, denoted by $VaR_{port}$. In this approach the dependence between asset price changes is implicitly taken into account in the creation of the portfolio by building the time-series for $\Delta P_{port}$.

**b) The risk factor approach**

In the risk factor approach, the distribution of price changes of each asset (more generally called "risk factors") is first estimated in order to compute the individual VaR of each asset, denoted by $VaR_1$ and $VaR_2$. The portfolio VaR is then computed by using an aggregation formula linking the individual VaR and the portfolio weights (more generally called "risk sensitivities"). A classical aggregation formula used in practice (see for example the RiskMetrics software developed by JP Morgan, 1995) computes the portfolio VaR denoted by $VaR_{port}^{agg}$ for a given $f$ as follows:

$$VaR_{port}^{agg} = \sqrt{x_1^2 \cdot (VaR_1)^2 + x_2^2 \cdot (VaR_2)^2 + 2 \cdot x_1 \cdot x_2 \cdot \rho_{12} \cdot VaR_1 \cdot VaR_2} \qquad (1)$$

This approach uses explicitly the correlation coefficient between asset price changes, denoted by $\rho_{12}$. This approach is also theoretically justified in the case of normality. If asset price changes are distributed according to a Gaussian distribution, then the two approaches to compute the VaR lead exactly to the same result.

## 3. Implied correlation

In this section we use the VaR calculation derived above in a reverse manner in order to estimate the implied correlation between asset price changes. By assuming that both approaches lead to the same VaR results, we compute the correlation coefficient in the aggregation formula that equates the two VaR. The correlation implied from VaR is given by:

$$\rho_{12} = \frac{(VaR_{port})^2 - x_1^2 \cdot (VaR_1)^2 - x_2^2 \cdot (VaR_2)^2}{2 \cdot x_1 \cdot x_2 \cdot VaR_1 \cdot VaR_2} \qquad (2)$$

If asset price changes are distributed according to a Gaussian distribution, then the same number should be found for the correlation coefficient. Especially, correlation should be autonomous from the following parameters:

- The probability used to compute the VaR: $p$,

- The weights used to build the portfolio: $x_1$ and $x_2$,

- The type of position: long or short,

- The frequency used to measure asset price changes: $f$ (under the i.i.d. assumption).

Under normality, the implied correlation does not depend on the parameters listed above and is simply equal to the classical Pearson correlation coefficient. Whether these set of propositions hold true is empirically tested in the next section.

## 4. Empirical results

The correlation implied from VaR is computed for a portfolio comprising the S&P 500 and FTSE 100 indexes. Data are closing index values over the time-period from January 1, 1995 to December 31, 2003. Two frequencies are used to measure price changes: daily and weekly. Different probability levels are used to compute the VaR: 80%, 95.45%, 98.46%, 99.23%, 99.62% and 99.81% (for daily VaR) and 75%, 92.31%, 96.15% and 98.08% (for weekly VaR). These probability levels correspond to average waiting time-periods of 1 week, 1 month, 1 quarter, 1 semester, 1 year and 2 years (for daily VaR) and 1 month, 1 quarter, 1 semester and 1 year (for weekly VaR). Different weights are chosen to build the portfolios: (25%, 75%), (50%, 50%) and (75%, 25%). Two types of position are used: long positions and short positions in both indexes. The individual and portfolio VaR are computed with historical distributions. Table 1 gives the correlation implied from daily VaR (Panel A) and from weekly VaR (Panel B). A graphical representation is also given in Figure 1.

Results indicate that the implied correlation tends to depend on the type of position (long or short), on the probability level used to compute the VaR and on the frequency used to measure price changes. However, the same pattern of implied correlation is obtained when portfolio weights vary.

Considering first results from daily VaR, implied correlation appears to be higher for long positions than for short positions and particularly for more extreme probabilities. For example, for an equally-weighted portfolio (50%, 50%), the implied correlation for long positions is higher than the Pearson correlation coefficient equal to 0.42 while it is systematically lower for short positions. The results indicate that the implied correlation for long and short positions tends to diverge as the probability level used to compute the VaR increases (that is when we look further in the distribution tails). For a probability level of 99.62% (corresponding to an average waiting time-period of 1 year), it is equal to 0.516 for the VaR on long positions and 0.261 for the VaR on short positions.

Implied correlation also depends on the frequency used to measure price changes. It is higher for weekly price changes than for daily price changes: 0.695 instead of 0.420 on average. Due to the lower number of weekly observations (divided by 5 compared to daily observations), the pattern of correlation implied from weekly VaR appears more erratic but once again, for high probability levels (or equivalently long average waiting time-periods), implied correlation is again much higher for long positions than for short positions. For example, for the equally-weighted portfolio and for a probability level of 99.08% (corresponding to an average waiting time-period of 1 year), the implied correlation is equal to 0.744 for the VaR on long positions and to 0.358 for the VaR on short positions. As for daily data, at the highest probability level, correlation for long positions (crashes) is double that of short positions (booms).

A similar pattern of implied correlation is found for different portfolio weights. For example, for the highest probability level, implied correlation from daily and weekly VaR of long positions is systematically higher than those of short positions.

## 5. Summary

The empirical results presented in the previous section can be summarized as the following stylized facts:

1) Implied correlation is not constant,

2) Implied correlation tends to be higher for long positions than for short positions,

3) Implied correlation tends to increase (decrease) with the probability level for long (short) positions,

4) Implied correlation tends to decrease overall with the frequency of price changes.

5) Implied correlation behaves in a similar way for different portfolio weights.

**Table 1. Correlation between US and UK index price changes implied from VaR.**

**Panel A. Correlation implied from daily VaR.**

| Probability (waiting period) | Portfolio weights | | | | | |
|---|---|---|---|---|---|---|
| | (25%, 75%) | | (50%, 50%) | | (75%, 25%) | |
| | Long | Short | Long | Short | Long | Short |
| 80% (1 week) | 0.407 | 0.410 | 0.402 | 0.374 | 0.340 | 0.310 |
| 95.45% (1 month) | 0.387 | 0.368 | 0.451 | 0.397 | 0.533 | 0.425 |
| 98.46% (1 quarter) | 0.351 | 0.403 | 0.412 | 0.403 | 0.430 | 0.571 |
| 99.23% (1 semester) | 0.466 | 0.571 | 0.597 | 0.291 | 0.572 | 0.158 |
| 99.62% (1 year) | 0.596 | 0.441 | 0.516 | 0.261 | 0.568 | 0.213 |
| 99.81% (2 years) | 0.511 | 0.138 | 0.484 | 0.209 | 0.345 | 0.214 |

**Panel B: Correlation implied from weekly VaR.**

| Probability (waiting period) | Portfolio weights | | | | | |
|---|---|---|---|---|---|---|
| | (25%, 75%) | | (50%, 50%) | | (75%, 25%) | |
| | Long | Short | Long | Short | Long | Short |
| 75% (1 month) | 0.564 | 0.651 | 0.506 | 0.675 | 0.077 | 0.781 |
| 92.31% (1 quarter) | 0.643 | 0.656 | 0.622 | 0.737 | 0.446 | 0.564 |
| 96.15% (1 semester) | 0.565 | 0.714 | 0.737 | 0.858 | 0.794 | 0.965 |
| 98.08% (1 year) | 0.684 | 0.491 | 0.744 | 0.358 | 0.682 | 0.391 |

*Note:* this table gives the correlation implied from daily VaR (Panel A) and from weekly VaR (Panel B) of long and short positions on portfolios composed of the S&P 500 and FTSE 100 indexes. Different probability levels (or equivalently average waiting time-periods) are used to compute the VaR. Different weights are chosen to build the portfolios.

**Figure 1. Correlation between US and UK index price changes implied from VaR.**

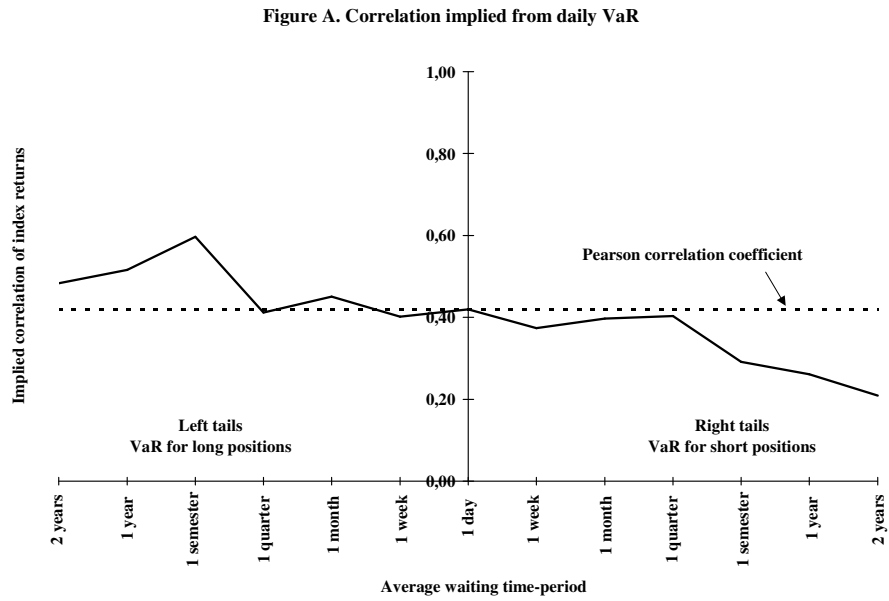

Figure A. Correlation implied from daily VaR

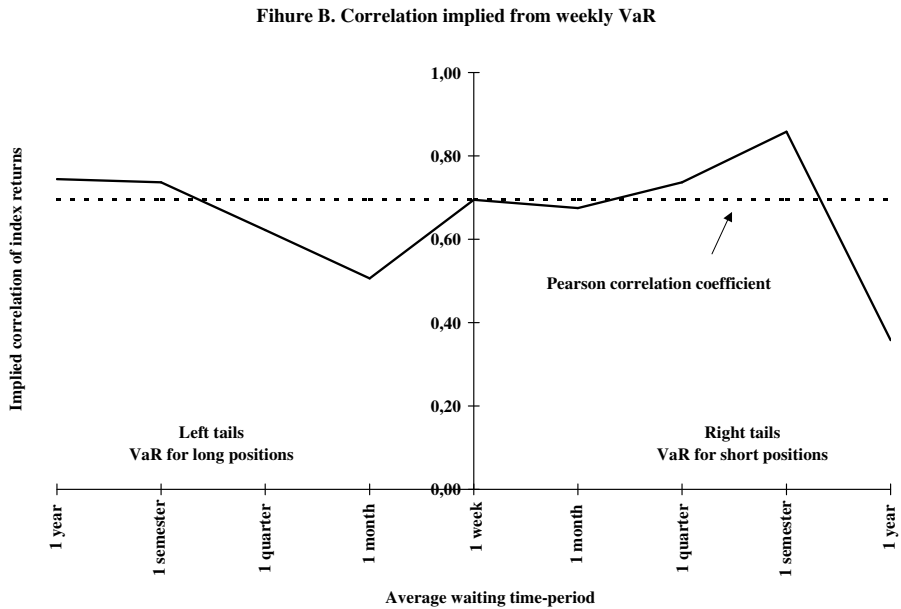

Fihure B. Correlation implied from weekly VaR

*Note:* this figure represents the correlation implied from daily VaR (Figure A) and from weekly VaR (Figure B) of long and short positions on an equally-weighted portfolio composed of the S&P 500 and FTSE 100 indexes for different average waiting time-periods (or equivalently probability levels) used to compute the VaR.